\documentclass[fleqn,espcrc2]{article}

\begin{document}

\title{Inferences from the dark sky:\\ Olbers' paradox revisited}
\date{}
\maketitle
\author{Mauro Arpino

{\em Civico Planetario "Ulrico Hoepli"

Corso Venezia 57, 20121 Milano, Italy.

email: arpino@tiscali.it} \\  \\

\and Fabio Scardigli\footnote{Corresponding author: Fabio Scardigli.\\ 
Address for all the correspondence: 
Via Europa 20 - 20097 San Donato Milanese - Milano, ITALY.}

{\em Institute for Theoretical Physics, University of Bern, 

Sidlerstrasse 5, 3012 Bern, Switzerland.

email: fabio@itp.unibe.ch}}

\begin{abstract}
The classical formulation of "Olbers' Paradox" consists 
in looking for an explanation of the fact that the sky at night is dark.
We use the experimental datum of the nocturnal darkness in order to put 
constraints on a newtonian cosmological model. We infer then that the 
Stellar System in such a model  
should have had an origin at a finite time in the past.\\ \\
PACS 98.80 - Cosmology. 
\end{abstract}

\section{Introduction}

The observation that the sky at night is dark seems to be in conflict 
with the idea of an infinite Universe. If the Universe is infinite
and contains an unlimited number of stars, our line of sight should meet 
a star in every direction we observe. Therefore, there should 
be no apparent gaps between stars on the celestial sphere. Moreover, repeating 
an argument originally attributed to de Cheseaux (1718-1751)~\cite{Che}, we note that, 
in an infinitely populated Universe, the number of stars contained in a spherical 
shell of radius $r$ is $4\pi n r^{2} dr$, where $n$ is the number of stars 
per unit volume. If $L$ is the absolute luminosity of a star, the intensity 
of light at a distance $r$ is $I=L/4\pi r^{2}$. Hence the generic shell 
produces a lighting (in the centre of the shell) equal to $nLdr$. 
Since there is an infinite number of shells, the lighting, in every point 
of the Universe, should be infinite!

But the sky at night is dark. This apparent contradiction has been named 
\textit{Olbers' Paradox}, following the publication in 1823 of H. W. Olbers' paper 
"On the Transparency of Space"~\cite{a}.
However, the first assertions about the conflict between the dimensions of 
the Universe and the nocturnal darkness can be traced back to Digges (1576),
Kepler (1610) and Halley (1721)~\cite{Harrison}.

Many explanations were given during the centuries in attempts to resolve the paradox.
Olbers suggested the presence of interstellar dust obscuring the farthest stars, 
but simple thermodynamical considerations by J.~Herschel (1848)~\cite{Herschel}
soon ruled out this explanation. Others invoked hypotheses that required a hierarchic 
distribution of stars ~\cite{Jaki}. In 1917 Shapley suggested the idea of an
"island-universe", that is a limited spatial extension of 
the stellar system, as a resolution of the paradox. 
After the discovery of the expansion of the Universe, the 
redshift was proposed as the fundamental mechanism necessary to "cut" the 
radiation coming from the farthest galaxies, hence assuring the darkness of night.

Harrison ~\cite{Harrison2} suggested that the paradox does not exist even in 
the context of a classical Universe, in the framework of which it was conceived, 
because there are too 
few stars and their lifetimes are too short. In other words the 
Universe does not contain enough energy to produce a bright sky. Nevertheless,
in current literature there are still many misunderstandings about the paradox. 
It is often explained using the redshift, which, on the contrary, does not 
have a relevant role ~\cite{Harrison3}. 
In fact detailed calculations show that in our Universe the luminosity of 
the sky is determined to order of magnitude by the lifetime of the galaxies,
and only affected by a factor of two by the expansion of the Universe 
(see ~\cite{Wesson&C}, which also gives a survey of the recent literature on the subject).

The classical formulation of Olbers' Paradox tries to explain 
the fact that the night is dark. In contrast, in the present paper, 
we use this observed fact to put some constraints on a simple 
euclidean-newtonian cosmological model. We then draw several necessary 
conclusions about the origin of the Universe (i.e. the Stellar System) 
at a finite time in the past.

\section{The cosmological model}

The model of space-time used here is the classical euclidean-newtonian one.
The Universe is a euclidean manifold $\Re \times \Re ^{3}$, with a euclidean 
metric on $\Re ^{3}$. The time we use is the absolute time of Newton. 
The generic event is singled out by orthogonal cartesian coordinates 
$(t, \vec{x})\in \Re \times \Re ^{3}$ .
Therefore space has an infinite extension and is isotropic, homogeneous and 
flat in every region and so it coincides with Newton's absolute space. 
The simultaneity we refer to is the newtonian absolute simultaneity.
The light signals propagate on euclidean straight lines and the speed of light 
is finite and equal to $c$.

In this model, all matter is in the form of stars. They are distributed in a homogeneous 
and isotropic way in the regions they fill. Therefore, there are no bunches of 
stars, such
as clusters, galaxies, etc. The number of stars per unit volume does not change 
with time. This hypothesis is equivalent to assuming that the lifetime of a 
"virtual star" is, \textit{a priori}, unlimited. 
From an astrophysical point of view, it means 
that a star does not have chemical, nuclear or gravitational evolution. 
Of course, based on current knowledge of stellar evolution, this is not true.
Yet, from the cosmological point of view, this idea is reasonable if we admit an 
uninterrupted sequence of stellar generations. For every star that dies, a new 
one comes up, so the density of "active" stars does not change over time.                
As we will see, the comparison of the model with the experimental evidence of 
the dark at night will allow us to decide between the hypothesis of an 
"eternal" stellar system (i.e. an infinite number of generations) or 
the hypothesis of a stellar system born at a finite time in the past.

Of course, the adopted model does not describe any of the main properties 
of the contemporary models (such as expansion, curvature, red shift, etc.). Yet, we have 
chosen this model for two reasons. The first is its metric simplicity, which will 
let us introduce the concept of "lookout limit" without mathematical complexity. 
The second reason is that this model coincides with the model of classical newtonian space-time, 
historically adopted by Halley, Olbers, Shapley, until the Einstein paper on cosmology 
(1917) ~\cite{Einstein}. 
Finally, we want to note that the closer the global curvature is to zero, 
the more the euclidean model approximates the curved space-time.

\section{Lookout limit and background radiation energy density}

Let us think of the stars as uniformly distributed in space. We then suppose that 
each star occupies an average volume $V$. Furthermore, each star has a section of area 
$\sigma$.
Thus, the probability that our line of sight meets a star surface is

\begin{equation}
p=\frac{\sigma}{V^{2/3}}.
\end{equation}
Our line of sight is certain to meet a star, if 
we observe a number $N$ of elementary volumes $V$, lined up in sequence, equal to

\begin{equation}
N=\frac{V^{2/3}}{\sigma}.
\end{equation}
This is true, of course, if the stars are distributed completely at random 
in their respective volumes.\\
The distance reached by the line of sight before meeting a star surface is 
therefore

\begin{equation}
\delta = V^{1/3} \frac{V^{2/3}}{\sigma}=\frac{V}{\sigma}.
\end{equation}
We call this distance \textit{lookout limit}, after Harrison. 
It is well known that this quantity 
coincides with the mean free path $\lambda$ of a photon ~\cite{Enge}; in fact 
$V=1/n$, where $n$ is the number of stars per unit volume and   

\begin{equation}
\lambda := \frac{1}{\sigma n} = \frac{V}{\sigma}=\delta.
\end{equation}
Following Harrison and Kelvin ~\cite{Harrison4}, we will derive an 
expression for the fraction of sky covered by stellar discs, and then a 
formula for the radiation energy density (at the point of the observer) 
due to the whole stellar system.\\ \\
Let us consider a spherical shell of radius $r$ and thickness $dr$.
Let $\Omega (r)$ be the solid angle \textit{free} of stellar discs up to the shell $r$.
Let $\Delta \Omega (r)$ be the solid angle intercepted by the stars present 
\textit{on} the shell $r$. The fraction of solid angle intercepted by stars on the shell 
$r$ is ~\cite{Enge}

\begin{equation}
\frac{\Delta \Omega}{\Omega} = \frac{total\: stellar\: area}{total\: shell\: area} = 
\frac{-n \sigma 4 \pi r^{2} dr}{4\pi r^{2}} = -n \sigma dr.
\end{equation}
After integration, we obtain an expression for the solid angle still 
\textit{free of stellar discs} after a radial path $r$

\begin{equation}
\Omega(r) = \Omega_{0} \, e^{-n\sigma r}
\end{equation}
where $\Omega_{0}\equiv 4\pi$.\\
The solid angle intercepted by the stellar surfaces after a path $r$ is therefore

\begin{equation}
\Omega_{0} - \Omega(r) = \Omega_{0}(1-e^{-n\sigma r}).
\end{equation}
Hence the \textit{fraction} of intercepted solid angle is

\begin{equation}
\alpha= \frac{\Omega_{0} - \Omega(r)}{\Omega_{0}} = 1 - e^{-n \sigma r}.
\end{equation}
$\alpha$ is the fraction of sky covered by stellar discs out to the radius $r$.
We note that if the stellar distribution is spatially unlimited, we can push 
$r \rightarrow \infty$, and then $\alpha \rightarrow 1$, i.e. the whole sky 
is covered by stars.\\
Also important is the relation

\begin{equation}
u(r)=4u^{\ast}(1-e^{-n \sigma r}),
\end{equation}
where $u(r)$ is the radiation energy density at the centre of a sphere of 
radius $r$, due to all the stars contained in the sphere, and $u^{*}$ is the 
radiation energy density at the surface of a star.\\
This relation can be derived as follows.\\
In a time $dt$ a star emits an energy $dE=Ldt$, where $L$ is the star's luminosity.
This energy spreads (in the time $dt$) on a volume $4 \sigma c dt$, where $4 \sigma$ 
is the surface area of a star with a section of area $\sigma$. 
Therefore we can write $dE=4u^{*}\sigma c dt$, where 
$u^{*}= L/(4\sigma c)$ is the radiation energy density at the surface of the star.
The radiated energy $dE$ travels until it reaches a distance $r$ 
from the star. Then, in the same time $dt$ as before, it spreads on a volume 
$4\pi r^{2} c dt$. Therefore the radiation energy density at a distance $r$ from
the source is

\begin{equation}
u=\frac{dE}{4\pi r^{2}cdt} = \frac{Ldt}{4\pi r^{2}cdt}= \frac{u^{*}\sigma}
{\pi r^{2}}.
\end{equation} 
Now, the number of stars on a shell of radius $r$ is 

\begin{equation}
4\pi r^{2} n dr,
\end{equation}
and the fraction of free sky, i.e. not covered by stellar discs at a distance 
$r$ from the observer, is (see eqs. (6), (8))

\begin{equation}
\frac{\Omega(r)}{\Omega_{0}} = e^{-n\sigma r}.
\end{equation}
Therefore the number of stars not obscured by other stars, i.e. visible at the centre 
of the shell, is

\begin{equation}
4 \pi r^{2} n dr e^{-n \sigma r},
\end{equation}
and the contribution to the radiation energy density at the centre of the shell 
due to the stars \textit{on} the shell is

\begin{equation}
du=4 \pi r^{2}n dr e^{-n\sigma r}\frac{u^{*} \sigma}{\pi r^{2}} =
4 u^{*} \sigma n e^{-n\sigma r}dr.
\end{equation}
Integrating this relation between $0$ and $r$, we obtain formula (9).\\
Hence $u$ represents the energy density of the background sky, or what is 
today called \textit{extragalactic background light}.
We note that $u=4u^{*}$ in every star distribution extending to 
$r\gg\lambda=1/n\sigma$. For an infinite spatial extension of the 
stellar system, the background sky energy density must be of the same order of that on 
the surface of a star. \\
The experimental condition of dark sky can be expressed in this context as

\begin{equation}
u\ll u^{*}.
\end{equation}
We note that the presence of the factor $e^{-n\sigma r}$ (which accounts for the 
mutual absorption of light by stars in the Universe) produces a lowering of lighting 
at any point in space from $\infty$ to $4 u^{*}$.

\section{Study of the model}

We are now able to examine all the cases suggested by the proposed model 
using the concept of lookout limit, having adopted absolute 
newtonian space and time, and having supposed that they both have an infinite 
extension.\\
For the stellar system, we have the two following possibilities:\\ 
\\
I) The distribution of the stellar system is spatially infinite.\\ 
\\
II) The distribution of the stellar system is spatially finite. We 
suppose then that it has a spherical symmetry with a radius $R$.\\ 
\\
In case I) we can distinguish two subcases:\\ 
\\
Ia) The stellar system has existed from an infinite time in the past.\\ 
\\
Ib) The stellar system was "turned on" (all the stars together and simultaneously)
at a time $t_{0}=-T$ in the past ($T>0$)(we take $t=0$ as the present time).\\ 
\\ 
In case Ia) it is evident that the sky at night must be luminous. In fact 
every line of sight should necessarily intercept, sooner or later, a star surface
($\alpha=1$). Therefore the celestial sphere must appear luminous, completely 
filled up by stellar discs, without dark spaces among stellar discs 
($u=4u^{*}$). Case Ia), a spatially infinite Universe, is the model adopted 
by Halley and Olbers, on the grounds of newtonian considerations. It is in 
evident conflict with observation, a conflict historically known as 
Olbers' Paradox.\\
\\
In case Ib), only the radiation of stars contained in
a sphere of radius $h=cT$ (cosmological horizon) can reach us. 
The radius of this sphere 
must be compared with the value of the lookout limit $\lambda$.
We can distinguish the following subcases.\\ 
\\
Ib1) If $\lambda>cT$ then the night sky could turn out to be "dark", that is 

\begin{equation}
u(cT)<u(\lambda)=4(1-\frac{1}{e})u^{*}.
\end{equation}
In fact, the lines of sight can extend outward to a distance of $r=\lambda$, 
beyond the "border" $r=cT$ of the stellar system.    
This situation is consistent with the observed evidence.\\ 
\\
Ib2) If $\lambda<cT$ then the sky at night must be "luminous", that is

\begin{equation}
u(cT)>4(1-\frac{1}{e})u^{*}.
\end{equation}
Of course this possibility is excluded by the experimental evidence.\\ 
\\
Let us now study case II). We have two sub-cases here as well:\\ 
\\
IIa) The stellar system has existed from an infinite time in the past 
($t_{0}=-\infty$);\\ 
\\
IIb) The stellar system was "turned on" at a time $t_{0}=-T$ in the past.\\ 
\\
In case IIa) we can compare $\lambda$ with the radius R of the stellar system:\\ 
\\
IIa1) If $\lambda >R$ then the night sky could be "dark", i.e. 
$u(R)<4(1-1/e)u^{*}$.\\ 
\\
IIa2) If $\lambda<R$ then the night sky should be "luminous", i.e. $u(R)>4(1-1/e)u^{*}$.\\ 
\\
The first case is in apparent agreement with the empirical evidence. 
Yet we observe that the assumptions "R finite" and "$t_{0}=-\infty$" create 
problems in the dynamical stability of the system. From the point of 
view of newtonian mechanics, a homogeneous sphere, with a finite radius
and a negligible angular momentum, made of particles interacting only by 
gravity, has to collapse inwards upon its geometrical centre in a finite 
time. For such dynamical reasons case IIa1) offers little of 
physical interest and can be ruled out.\\
\\
In case IIb), stars were born at a time $t_{0}=-T$ in the past. 
The quantities to be compared are now the lookout limit $\lambda$, 
the radius $R$ of the 
stellar system, and the horizon $h=cT$ of the observable spherical region 
of the stellar system. We have therefore $6$ ($=3!$) different cases. Their 
analysis is summarized by the following table.\\ 

IIb1) $cT \quad < \quad  \lambda \quad < \quad R \quad \Rightarrow \quad $ 
(dark sky: $u(cT)<4(1-1/e)u^{*}$) \\ \\

IIb2) $\lambda \quad < \quad cT \quad < \quad R \quad \Rightarrow \quad $ 
(luminous sky: $u(cT)>4(1-1/e)u^{*}$) \\ \\

IIb3) $\lambda \quad < \quad R \quad < \quad cT \quad \Rightarrow \quad $ 
(luminous sky: $u(R)>4(1-1/e)u^{*}$) \\ \\

IIb4) $R \quad < \quad  \lambda \quad < \quad cT \quad \Rightarrow \quad $ 
(dark sky: $u(R)<4(1-1/e)u^{*}$) \\ \\

IIb5) $cT \quad < \quad  R \quad < \quad \lambda \quad \Rightarrow \quad $ 
(dark sky: $u(cT)<4(1-1/e)u^{*}$) \\ \\

IIb6) $R \quad < \quad  cT \quad < \quad \lambda \quad \Rightarrow \quad $ 
(dark sky: $u(R)<4(1-1/e)u^{*}$). \\
\\
We should point out that in cases IIb) also, with $R<\infty, \: t_{0}=-T$,
there could be problems of dynamical instability. We can say that the collapse 
of the spherical stellar system is unavoidable; nevertheless we can think 
that the collapse is not observable if we suppose that it is very slow 
compared with the cosmic time scale (T) (speed of collapse $\ll c$).\\
It is interesting to note that Newton himself investigated the problem of the 
dynamical stability of a finite spherical system of stars. He knew that such 
a system should have collapsed after a finite time inwards upon its centre of mass. 
But, strangely enough, he avoided calculating the time for gravitational collapse.
Only in 1902, did Lord Kelvin complete the first cosmological calculus on the 
collapse time of the whole stellar system ~\cite{Harrison4}.
He found that the collapse time does not depend on the initial size of the 
system. Instead, it depends only on its initial density

\begin{equation}
t_{k}= \left(\frac{3\pi}{32 G \rho_{0}}\right)^{1/2}.
\end{equation}
Assuming an average interstellar distance of 1 pc and a mean mass of 1 
solar mass, Kelvin obtained a time for the collapse equal to $20 \times 10^{6}$
years. Yet, Kelvin used classical arguments: he ignored the relativistic limit 
of the speed of light and therefore he obtained such a short 
collapse time that it was not realistic if compared with the cosmic time scale. 
If we take into account the limiting speed of light $c$, then the collapse 
time cannot be less than the radius of the universe divided by $c$, because
the various regions will collapse, at the fastest, with a speed equal to $c$.

\section{Conclusions}

The comparison of the cases presented in the above section with the 
experimental evidence of the dark sky at night, allows us to 
eliminate cases Ia, Ib2, IIa2, IIb2, IIb3. They produce a luminous 
sky at night. Only cases Ib1, IIb1, IIb4, IIb5, IIb6 are physically 
admissible, because they are consistent with the experimental evidence.
What property do they share?

The main consideration is that all these cases imply \textit{the birth of
the Stellar System at a finite time in the past}. We note that our model, 
together with the evidence of the dark sky, is not able to discriminate 
between a Universe with a finite radius and one with an infinite radius.       

Therefore the conclusion that can be drawn from our model is 
that the experimental evidence of the dark sky implies that \textit{the 
Stellar System was born at a finite time in the past}.
This important observation was not formulated by Olbers or others (Halley, 
Kelvin, etc.). Olbers would have been able to infer, from the
observed fact of a dark night sky, the birth of the cosmos at a finite time in the past.
This could happen because the cosmological model adopted by Olbers (and 
more or less explicitly by all the scientists until Einstein) is the 
newtonian one described in section 2. The lookout limit introduced by 
Harrison is a useful concept which has allowed us to obtain a deeper and more
effective understanding of the model.

We want to emphasize that other authors have focused particularly on the 
luminous lifetime of single stars. But this has prevented 
them from reaching the relevant conclusions about the lifetime of 
the \textit{whole} stellar system, i.e. the Universe. The main hypothesis that 
allowed us to obtain information about the lifetime of the Universe (the Stellar 
System) is 
the one of the "virtual star", i.e. a fictitious luminous source that, in 
principle, has an infinite life. Successive generations of real stars 
can be modelled as a single virtual star without an \textit{a priori} fixed 
lifetime. The check with the observational evidence ($u\ll u^{*}$) 
determines, at this point, the lifetime of the virtual star, i.e. of the whole 
Stellar System. In other words, we could say that classical physics (without hypotheses
on nuclear reactions and expansion of the Universe) is able to explain the 
nocturnal darkness of the sky only by requiring a finite lifetime of the Stellar System.

Finally we want to observe that several authors (e.g. Weinberg) state that 
in a Big Bang cosmology there is no paradox because the contribution of 
the various stellar generations is cut at a finite time in the past 
~\cite{Weinberg}. This is not always true from our point of view, because we
have seen that we can have a "Big Bang cosmology" (in the newtonian sense, 
namely that the stellar system was "turned on" in the past), 
which nonetheless produces a luminous sky (cases Ib2, IIb2, IIb3).
In conclusion, Olbers would not have been able to infer the expansion of the 
Universe from the observation of the dark sky at night, as several authors 
~\cite{Bondi} seem to suggest. On the contrary, Olbers and other 
authors (from the 18th to the early 20th centuries) would have been able to assert that 
the Universe has had a temporal origin in the past.\\ \\

\large
\textbf{Acknowledgements} \\ \\
\normalsize
The authors wish to thank the anonymous referee for useful comments, 
T.J. Lindsey for style suggestions, and the nice
Irish Pub "Crazy Patrick", where this paper was conceived.

\end{document}